\documentclass[10pt,twocolumn,superscriptaddress,aps,prl,showpcas,preprintnumbers,amsmath,amssymb,floatfix]{revtex4}
\usepackage[latin9]{inputenc}
\usepackage{color}
\usepackage{graphicx}
\usepackage{esint}
\usepackage[unicode=true, bookmarks=false, breaklinks=false,pdfborder={0 0 1},backref=false,colorlinks=true]{hyperref}

\makeatletter
\@ifundefined{textcolor}{}
{
 \definecolor{BLACK}{gray}{0}
 \definecolor{WHITE}{gray}{1}
 \definecolor{RED}{rgb}{1,0,0}
 \definecolor{GREEN}{rgb}{0,1,0}
 \definecolor{BLUE}{rgb}{0,0,1}
 \definecolor{CYAN}{cmyk}{1,0,0,0}
 \definecolor{MAGENTA}{cmyk}{0,1,0,0}
 \definecolor{YELLOW}{cmyk}{0,0,1,0}
}

\usepackage{graphicx}
\usepackage[english]{babel}
\usepackage{amsmath}
\usepackage{amssymb}

\newcommand{\bQ}{{\mathbf \Xi}}
\newcommand{\bP}{{\mathbf \Pi}}
\newcommand{\cP}{{\mathbf P}}
\newcommand{\D}{{\mathbf \Delta}}

\newcommand{\bC}{{\mathbf C}}
\newcommand{\bG}{{\mathbf G}}

\newcommand{\bT}{{\mathbf T}}
\newcommand{\bV}{{\mathbf V}}
\newcommand{\bI}{{\mathbf I}}

\newcommand{\tr}{\mbox{\mbox{Tr}}}
\newcommand{\bx}{{\mathbf x}}

\newcommand{\vb}[1]{\mathbf{#1}}
\newcommand\sups[1]{^{\hbox{\scriptsize{#1}}}}

\begin{document}
\title{Entanglement entropy of dispersive media \\
from thermodynamic entropy in one higher dimension}
\author{M. F. Maghrebi}
\email[Corresponding author: ]{magrebi@umd.edu}
\affiliation{Joint Quantum Institute, NIST/University of Maryland, College Park, Maryland 20742, USA}
\affiliation{Joint Center for Quantum Information and Computer Science, NIST/University of Maryland, College Park, Maryland 20742, USA}
\author{M. T. H. Reid}
\affiliation{Department of Mathematics, Massachusetts Institute of Technology, Cambridge, MA 02139, USA}

\begin{abstract}
A dispersive medium becomes entangled with zero-point fluctuations in the vacuum. We consider an arbitrary array
of material bodies weakly interacting with a quantum field
and compute the quantum mutual information between them.
It is shown that the mutual information in $D$ dimensions can be mapped to classical thermodynamic entropy in $D+1$ dimensions.
As a specific example, we compute the mutual information both analytically and numerically for a range of separation distances between two bodies in $D=2$ dimensions and find a logarithmic correction to the area law at short
separations.  A key advantage of our method is that it
allows the strong subadditivity property to be easily verified.
\end{abstract}

\maketitle

Entanglement as a purely quantum phenomenon has become a central
subject in many fields, ranging from high-energy physics to many-body
systems and quantum information. In particular, the characteristic
entropy associated with entanglement has proved a powerful tool for
describing the quantum nature of a system \cite{Vidal03,Levin96,Kitaev96,Calabrese04} as well
as for quantifying resources
available for quantum computation
\cite{Bennett96}. Inspired by formal similarities
with thermodynamics, parallels between quantum and classical notions
of entropy have been suggested in the literature \cite{Popescu97}.

While many studies of entanglement entropy have considered geometrical
regions in vacuum \cite{Calabrese04} or domains in many-body systems \cite{Amico08}, rapid
advances in quantum computation and atomic and optical physics have
prompted studies of entanglement between atoms and
light \cite{Lambert04, Hur08}.
Although most studies have
addressed the case of one or a few bodies, recent attention has turned
to the problem of quantifying the entanglement of a medium interacting
with a quantum field, both described by (infinitely) many modes.
In particular,
Ref.~\cite{Klich12} took key first steps in this
direction, but simple conceptual pictures and efficient computational
tools have remained elusive.

In this letter, we consider two or more dispersive media
in the presence of a fluctuating field and compute their quantum mutual
information, a characteristic measure of quantum correlations between parts
of a system, and closely related to the entanglement entropy. Our model is a simplified electromagnetism in
which a free scalar field couples weakly to ``dielectric'' bodies.
We relate the mutual information in $D$ dimensions to classical
thermodynamic entropy in $D+1$ dimensions, where the dispersive media become planar disk-like regions (Fig.~\ref{Fig. DtoD+1}).
Inspired by an electrostatic analogy, our formalism relates
quantum information to classical scattering theory, thus
allowing practical computations by any number of techniques
including multipole expansion or numerical methods.
As an example, we compute the mutual information between
two bodies at all separations in $D=2$ dimensions and observe
area-law-violating logarithmic behavior at short separations.
\begin{figure}[t]
\centering
\includegraphics[width=9.5cm]{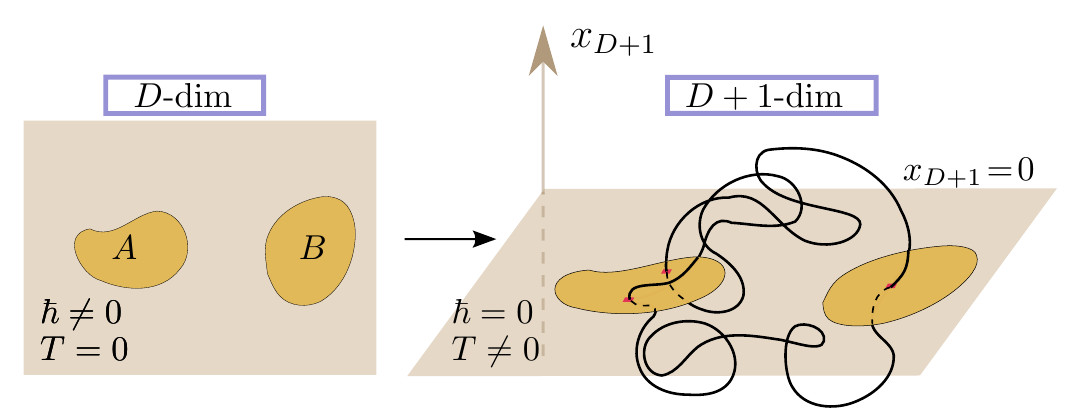}
  \caption{Two dispersive media in $D$ dimensions coupled to a quantum field in vacuum.
The mutual information between the two is related to the thermodynamic entropy of fluctuating fields in $D+1$ dimensions. The material bodies $A,B$ are reinterpreted as subregions in the plane $x_{D+1}=0$ in $(D+1)$-dimensional space, and impose a boundary condition on the fluctuating fields. This analogy finds a geometrical picture in terms of phantom chain polymers intersecting both regions.}
\label{Fig. DtoD+1}
\end{figure}

\emph{Model.}
Our starting point is the action
describing a free scalar field in $D$ spatial dimensions
in the presence of dispersive material bodies,
\begin{equation}\label{Eq. Action of phi}
   I_{\rm di}[\phi]=\int_{0}^{\infty}\!\!\frac{d\omega}{2\pi} \!\int \!\!d\bx\, \phi^*_\omega(\bx)\left[\epsilon(\omega, \bx)\,\omega^2+\nabla^2\right] \phi_\omega(\bx),
\end{equation}
where $\epsilon=1+4\pi\chi$ is a frequency- and position-dependent
dielectric function. This model offers a simplified theoretical framework
for exploring quantum electrodynamics and has proven useful in studies of Casimir~\cite{Kenneth06,Maghrebi14} and
entanglement~\cite{Klich12} phenomena. Specifically, the limit $\epsilon \to 1$ is the basis of a vast literature on entanglement entropy. We take the susceptibility
$\chi$ to be nonvanishing in a region $\Omega$ (the
material bodies), within which it assumes the spatially constant form
$\chi(\omega)=\frac{\omega_p^2}{ \omega_0^2-\omega^2}$
with $\omega_0$ and $\omega_p$ the resonant and plasma frequencies of the dispersive material, respectively.
Equation~(\ref{Eq. Action of phi}) may be viewed
as an \textit{effective} action arising upon integrating out the
matter degrees of freedom
from a parent action describing both vacuum and matter
fields~\cite{Klich12}, which (upon Wick rotation, $\omega\to i\xi$) reads
\begin{align}\label{Eq. Action of phi and psi}
   &I_E[\phi, \psi]=\int_0^\infty \frac{d\xi}{2\pi} \Big\{\int  d\bx \, \phi_\xi(\bx)\left(\xi^2-\nabla^2 \right) \phi_\xi(\bx)\, \nonumber \\
    &+ \int_{\Omega} \!\!d\bx  \Big[\frac{\xi^2+\omega_0^2}{4\pi} \,  \psi^2_\xi(\bx)+ 2\omega_p \xi\,\phi_\xi(\bx) \psi_\xi(\bx) \Big]\Big\},
\end{align}
where $\psi$ represents the degrees of freedom within the material bodies.
Note that the action in the first line is that of a free scalar
field, while the first term in the second line describes harmonic
oscillator modes localized in $\Omega$, and the last term is assumed to be of the form $\psi d\phi/dt$ \footnote{Here we have closely followed Klich \cite{Klich12}; see also his discussion
on the last term in Eq.~(\ref{Eq. Action of phi and psi}).}. Integrating
out the field $\psi$ from Eq.~(\ref{Eq. Action of phi and psi})
recovers the action (\ref{Eq. Action of phi}).  On the other hand, we are interested in finding an effective action for the matter field $\psi$.
Integrating out the scalar field $\phi$ yields
\begin{align}\label{Eq. Action of psi}
   I_{\rm eff}&[\psi]
  =
   \int_0^\infty \frac{d\xi}{2\pi}
     \Big\{\int_{\Omega} d\bx  \frac{\xi^2+\omega_0^2}{4\pi} \, \psi^2_\xi(\bx)
\nonumber \\
   &+\omega_p^2\xi^2\int\!\!\!\int_{\Omega} \!\!d\bx \, d\bx' \,
     \psi_\xi(\bx) G_\xi(\bx,\bx') \psi_\xi(\bx')\Big\},
 \end{align}
where $G_\xi=(\xi^2-\nabla^2)^{-1}$ is the Green's function of the
Euclidean Helmholtz operator.
The above effective action will allow us to study the entanglement between different parts of medium (or between two media). To this end, we briefly introduce the main tools to compute the entanglement entropy.

\textit{Entanglement from covariance matrices.}
For a mixed state $\rho$, the von Neumann entropy, a widely used measure of entanglement, is defined as $-\tr[\rho \log \rho]$.
Let us consider the local field $\{ \psi(t,\bx)\}$ and its conjugate momentum $\{\pi (t,\bx)\}$ describing the local degrees of freedom in the region $\Omega$.
For a quadratic action, the ground state is a Gaussian functional of $\{ \psi\}$ (or $\{\pi\}$). Therefore, the quantum state can be fully represented by equal-time two-point functions $\bQ(\bx,\bx') =\langle \psi(t,\bx) \psi(t,\bx')\rangle$ and $\bP(\bx,\bx') =\langle \pi(t,\bx) \pi(t,\bx')\rangle$. The von Neumann entropy is directly related to the two-point functions via ${\cal S}_\Omega=
\tr\left[\left(\D_\Omega+\bI\right)\log\left(\D_\Omega+\bI\right)-\D_\Omega\log\D_\Omega\right]$ where $\bI$ is the identity matrix, and $\D_\Omega \equiv\sqrt{\bP \,\bQ }-\bI/2$ is the covariance matrix \cite{Botero03,Peschel09} with the subscript $\Omega$ emphasizing its zero support outside $\Omega$.
We find it more convenient to recast the von Neumann entropy as
\begin{align}\label{Eq. id}
  &{\cal S}_\Omega= \int_0^{1} d\lambda \,{s}_\Omega(\lambda) \nonumber \\
  \mbox{with} \quad &{s}_\Omega(\lambda)=\tr \log\left[\lambda^{-1} \D_\Omega+\bI\right],
\end{align}
where $\lambda\in\left[0,1\right]$ is an auxiliary parameter. The last expression is reminiscent of a one-loop effective action. In fact, we shall see that, by using the above identity, the entanglement entropy finds a convenient path-integral form.

Next we compute the covariance matrix $\D_\Omega$.
The correlation functions in imaginary frequency are given by
\(
    \bQ(\bx,\bx')=\int\frac{d\xi}{2\pi} \, \langle \psi_{ \xi}^*(\bx)  \psi_{ \xi}(\bx') \rangle
\)
and
\(
      \bP(\bx,\bx')=-\int\frac{d\xi}{2\pi}\, \xi^2\, \langle \psi_{ \xi}^*(\bx)  \psi_{ \xi}(\bx') \rangle \nonumber
\) \footnote{The two-point function of conjugate momenta
should be regularized by subtracting an unimportant divergent term.};
the two-point function in the integrands can be derived from Eq.~(\ref{Eq. Action of psi}).
In general, this is a difficult task which requires knowledge of the full
(off-shell) $\bT$-matrix---an object that appears in the Lippmann-Schwinger equation---of the material bodies in position space \cite{Kenneth06}. Nevertheless, we consider a weak-coupling limit in
which $\omega_{0,p} L\ll 1$ with $L$ the linear size of the media. In other words, the bodies are sufficiently small that
their coupling to the background scalar field [the second line in Eq.~(\ref{Eq. Action of psi})] can be treated perturbatively, a condition
well satisfied for sub-micron-scale bodies with typical plasma and
resonant frequencies. One can then expand the correlation functions perturbatively:
$\bQ\approx(4\pi/\omega_0)(\bI/2 +\delta \bQ)$ and
$\bP\approx(\omega_0/4\pi) (\bI/2+\delta \bP)$.
In the absence of interaction with
the scalar field, one recovers the standard variances of the field
and its conjugate momentum as those of a harmonic
oscillator, whereupon $\D_\Omega \to 0$ and the entanglement vanishes.
To leading order, we have $\D_\Omega \approx (\delta\bQ+\delta\bP)/{2}$.
Within this approximation, we find that $\D_\Omega= \cP_\Omega \D\cP_\Omega$, with $\cP_\Omega$ the projection operator onto the spatial domain of $\Omega$, and
\begin{equation}
  \label{Eq. D-matrix from G0}
   \D(\bx,\bx')= \omega_c \bG_0^{(D+1)}(\bx,\bx'),
\end{equation}
where we have put $\omega_c\equiv2\pi\omega_p^2/\omega_0$. This equation relates the covariance matrix $\D$ to the Green's function for the Laplacian in $D+1$-dimensions, $\bG_0^{(D+1)}(\bx,\bx')\sim 1/|\bx-\bx'|^{D-1}$; see the
{Supplemental Material} \cite{supp}.
It is perhaps surprising that $\D(\bx,\bx')$ depends only on
$\bx$ and $\bx'$, and not the specific geometry of $\Omega$; however, this is an approximation, and the geometric dependence appears in higher orders in $\omega_{0,p} L$.

Now we are in a position to write Eq.~(\ref{Eq. id}) in a familiar form. First note that, with $\D_\Omega= \cP_\Omega \D\cP_\Omega$, the second line of Eq.~(\ref{Eq. id}) can be written as $\tr \log\left[\left(\lambda^{-1} \cP_\Omega +\D^{-1}\right)/\D^{-1}\right]$. The inverse of $\D$ finds an awkward nonlocal form in $D$ dimensions, while it is simply proportional to the Laplacian (hence, local) operator in $D+1$ dimensions.
Therefore, we can cast the above expression as a functional integral over a real-valued field $\theta(\bx,x_{D+1})$ living in $D+1$ dimensions [$(\bx,x_{D+1}) \in {\mathbb R}^{D+1}$],
\begin{equation}\nonumber
{s}_\Omega(\lambda)=-2\log \frac{\int D\theta \, \exp\left[-\omega_c^{-1}{\int_{D+1}\left(\nabla\theta\right)^2- \lambda^{-1} \int_\Omega \theta^2}\right]}{\int D\theta \, \exp\left[-\omega_c^{-1}{\int_{D+1}\left(\nabla\theta\right)^2}\right]},
\end{equation}
which is easily seen to reproduce Eq.~(\ref{Eq. id}) upon functional integration.
We emphasize that $\theta$ is an auxiliary field, and is not related to the original fields in the model.
The exponent in the numerator inside the logarithm can be interpreted as a Hamiltonian which is a sum of the free and potential terms: The first term is the usual gradient term in $D+1$ dimensions, while the latter is the local potential felt only in the $D$-dimensional region $\Omega$ where $x_{D+1}=0$ (see Fig.~\ref{Fig. DtoD+1} with $\Omega=A \cup B$). Thus the terms inside the logarithm take the form of a partition function of the classical field $\theta$ at a finite temperature; we choose $k_B T=1$ for convenience. The above expression then characterizes the change of the \emph{classical free energy} in the presence of the potential term, that is $-F_\Omega(\lambda)= \log\left\langle \exp\left[-\lambda^{-1} \int_\Omega \theta^2\right] \right\rangle_0$ where $\langle \cdots \rangle_0$ denotes the average with respect to the Gaussian fluctuations of the gradient term.

We will argue later that the change of the free energy is purely entropic, and make the connection to the thermodynamic entropy. For now, we point out another simplification: $\lambda$ can be scaled by $\omega_c^{-1}$ to eliminate the latter in the gradient term. This gives rise to an overall factor of $\omega_c$ to the integral in Eq.~(\ref{Eq. id}) and changes its upper bound to $\omega_c^{-1},$ which we extend to infinity as $\omega_c L\ll 1$. The von Neumann entropy is then obtained from the free energy of the thermodynamic model described above as
\begin{equation}\label{Eq. S from F}
   {\cal S}_\Omega=2\omega_c \int_{0}^{\infty} \!\!\!d\lambda \,\, F_\Omega(\lambda).
\end{equation}
The degrees of freedom in $\Omega$ are uncountably infinite and thus the entropy depends on a UV cutoff \cite{Eisert10}; such dependence may exhibit some form of universality in one dimension \cite{Callan94,Holzhey94,Calabrese04}. We, on the other hand, consider the quantum mutual information (QMI) between two subsystems $A$ and $B$ defined as ${\cal I}_{A,B}={\cal S}_A+{\cal S}_B-{\cal S}_{A \cup B}$, which is inherently cutoff-independent, and measures quantum correlations between $A$ and $B$. Equation (\ref{Eq. S from F}) then relates the QMI to the change of the free energy with the local potential turned on in one or both regions, $\Delta F_{A,B}(\lambda)=F_{A\cup B}(\lambda)-F_A(\lambda)-F_B(\lambda)$,
\begin{equation}\label{Eq. I from F}
   {\cal I}_{A,B} =-2\omega_c \int_{0}^{\infty} \!\!\!d\lambda \,\, \Delta F_{A,B}(\lambda).
\end{equation}
The latter, however, is a familiar quantity in the context of thermal Casimir effect that gives rise to an interaction between two objects due to thermal, rather than quantum, fluctuations in the space between them \cite{Kardar91}. The relation in Eq.~(\ref{Eq. I from F}) enables us to use the powerful tools available to study the (thermal or quantum) Casimir effect. It is well-known that one can compute the Casimir energy between two objects from the knowledge of their classical scattering matrices, i.e. their response to incoming classical waves \cite{Kenneth06,Emig08}. In the case of the thermal Casimir effect, one should compute generalized \emph{capacitance} matrices, or the multipole (monopole, dipole, etc.) response of the objects to an external electrostatic potential. The interested reader can find more details in Refs. \cite{Kenneth06,Emig08}; we quote the final expression for $\Delta F_{A,B}(\lambda)$,
 \begin{equation}\label{Eq. CGCG formula}
   \Delta F_{A,B}(\lambda)=\frac{1}{2}\,\tr_{(D+1)}
              \log\left[\bI- \bC_A(\lambda) \bG_0 \bC_B(\lambda) \bG_0\right],
 \end{equation}
where $\bG_0$ and $\bC_i$ denote the $D+1$-dimensional Green's function and capacitance matrix elements, respectively. To understand the $\lambda$-dependent $\bC_A$, we note that, in the absence of fluctuations, the field $\theta$ satisfies $\left[-\nabla^2+V_A\right]\,\theta=0$; the derivatives in this equation act in $D+1$ dimensions, and the potential is defined as $V_A(\bx, x_{D+1})=\lambda^{-1}\delta(x_{D+1})$ when $\bx\in A$, and 0 otherwise.
As the potential is confined to the plane $x_{D+1}=0$, its effect is simply to impose the boundary condition [$\zeta \equiv x_{D+1}$]
\begin{equation}\label{Eq. B.C.}
   -\partial_\zeta \theta{\Big|}_{\, 0^-}^{\,0^+}+{\lambda}^{-1}\,\theta(\bx, \zeta=0)=0, \qquad \bx \in A,
\end{equation}
on the otherwise free electrostatic field satisfying $\nabla^2 \theta=0$. For example, $\lambda=0$ corresponds to a perfect conductor, while in the limit $\lambda\to \infty$ the object is completely transparent. The generalized capacitance matrix elements in $\bC_A(\lambda)$ then characterize the electrostatic response of the object due to the $\lambda$-dependent boundary condition in Eq.~(\ref{Eq. B.C.}). For the sake of completeness, we mention that the capacitance can be expressed as $\bC_A=\bV_A\frac{1}{\bI + \bG_0 \bV_A}$, where $\bV_A$ is the operator-valued potential corresponding to $V_A$. This equation is reminiscent of the Lippmann-Schwinger equation, which is yet another route to scattering theory.

In short, Eqs. (\ref{Eq. I from F},\ref{Eq. CGCG formula}) are the central results of this paper, and relate quantum mutual information in $D$ dimensions to classical (generalized) capacitance elements in $D+1$ dimensions, which allows exploitation of myriad analytical and numerical methods for computing $\bC$-matrix elements. Next we consider specific examples to showcase the efficiency of our method.

We start with two material bodies (Fig.~\ref{Fig. DtoD+1}) separated by a large distance $d$ compared to their sizes (yet small compared to $\omega_c^{-1}$).
The $\bC$-matrix elements may be computed analytically for simple shapes via techniques reminiscent of electrostatics.
For large $d$, we can keep only the monopole coefficient in Eq.~(\ref{Eq. CGCG formula}), combined with Eq.~(\ref{Eq. I from F}), to find
\begin{equation}
{\cal I}_{A,B}\approx
  \frac{\omega_c}{A_{D+1}\, d^{2(D-1)}}\int_0^\infty d\lambda \,
   C^0_A(\lambda) C^0_B(\lambda),
\end{equation}
where $A_D=[\Omega_D (D-2)]^2$ with $\Omega_D$ the area of the
unit $D$-sphere, and $C^0$ is the monopole element of the generalized capacitance matrix. Cardy \cite{Cardy13} also finds the same power-law and a similar expression for the QMI in terms of the
capacitance for a scalar field theory, albeit in the absence of matter; see also \cite{Casini09,Shiba12}.
This suggests that we can view material bodies coupled to a free field as
a \emph{probe} of quantum information shared between different regions in space.

As a specific example, we consider two disc-shaped bodies in $D=2$ dimensions of radii $R_{A,B}$. For a single disc of radius $R_i$, one computes
$C^0_i(\lambda)=\frac{R_i}{4\lambda /R_i +\pi/2}$; see the {Supplemental Material}. The QMI is then given by
 \begin{align}
  \label{Eq. Two disks at large sep}
  {\cal I}_{A,B}&= \frac{\omega_c R_A^2 R_B^2 \log{R_B}/{R_A}}
           {32 \pi^3 (R_B-R_A) d^2} \,\,
      \xrightarrow{R_A \to R_B=R} \,\,\frac{\omega_c R^3}{32\pi^3 d^2}.
  \end{align}
Note that, even at large separations, the mutual information depends
nontrivially on the radii, reflecting the difficulty of computations in general.

At short separations $d \lesssim R$, on the other hand, the multipole
expansion becomes inconvenient due to the need to retain large numbers
of multipoles.
\begin{figure}[t]
\centering
\def\svgwidth{10.5cm}
\resizebox{0.5\textwidth}{!}{\includegraphics{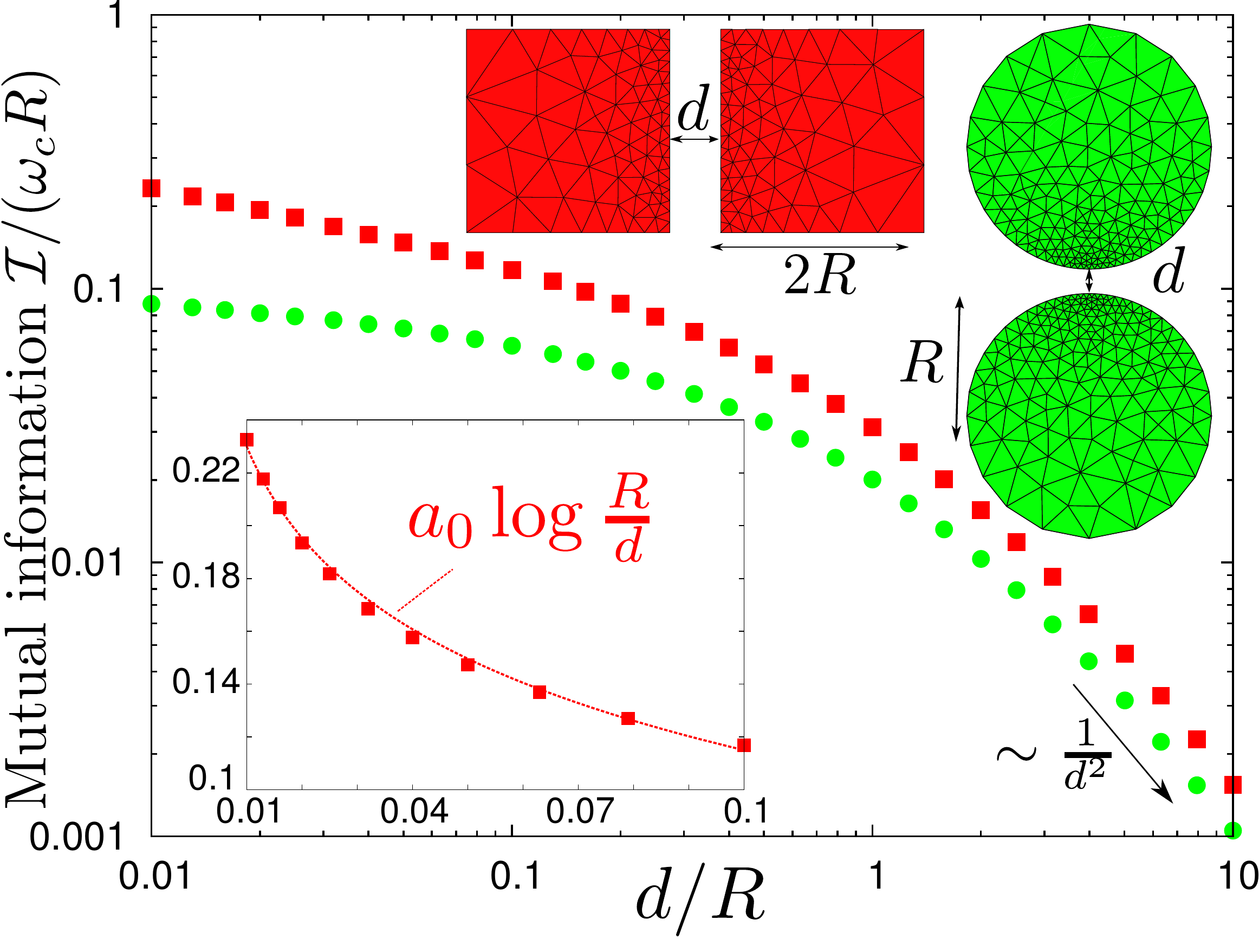}}
\caption{Quantum mutual information (QMI) vs. separation for circular
(green circles) or square (red squares) material bodies
in $D=2$ dimensions, as computed from Eqs.~(\ref{Eq. I from F},\ref{Eq. CGCG formula}) using the numerical method
outlined in the text with the bodies discretized into
small triangles (upper insets). At large separations, the
QMI obeys a power law, while at short separations the
square--square data suggest a logarithmic dependence on $d$.
The lower inset shows an enlarged view of the short-distance
data for squares, together with $a_0=0.050.$}
\end{figure}
In this case it is useful to note that, while the $\bC$
operator itself is unwieldy to describe in real space, the
\textit{inverse} of this operator has a simple real-space
representation~\cite{Rahi09}, namely
$ \bC^{-1}(\bx, \bx^\prime) =
   \lambda\delta(\bx-\bx^\prime) + \bG_0(\bx, \bx^\prime).
$
It is then tempting to discretize regions $A$ and $B$
into small finite elements (Fig.~2) and introduce a
basis of $N$ expansion functions $\{b_\alpha(\bx)\}$
localized on these elements; in this basis, we easily compute
matrix elements of $\bC^{-1}$ directly
according to
$ C^{-1}_{\alpha\beta}(\lambda) \equiv
   \big \langle b_\alpha
   \big| \bC^{-1}(\lambda)
   \big| b_\beta \big \rangle
$,
then simply invert the matrix $\bC^{-1}$ numerically to yield the
$\bC$ matrix for arbitrary $\lambda$ (with $\bG$ computed
similarly). The operator products and the trace in
Eq.~(\ref{Eq. CGCG formula}) become $(N\times N)$-dimensional
matrix operations and the $\lambda$ integral is evaluated by
numerical quadrature.
As two examples, Fig.~2 plots the predictions of Eq.~(\ref{Eq. CGCG formula})
for disc-shaped and square-shaped bodies evaluated numerically
for a range of separations $d$.
In the long-distance regime, the QMI follows the power-law in
Eq.~(\ref{Eq. Two disks at large sep}) and increases monotonically as the separation becomes shorter. For small separations $d\ll R$, the
numerical data suggests a logarithmic dependence on $d$,
\begin{equation}
    {\cal I}_{A,B}\sim \omega_c R \, \log\frac{R}{d}.
\end{equation}
Note that the mutual information is proportional to the ``area'' of the common boundary of the two media up to a logarithm, in this case the area being the common edge of the two-dimensional regions. The area law can be proved under certain conditions \cite{Hastings2007,Wolf08,Masanes09}; for a review, see Ref.~\cite{Eisert10}. Logarithmic violations of the area law also appear in several contexts \cite{Callan94,Holzhey94,Wolf06, Klich06,Swingle10,Leschke13}.

\emph{Mutual information from thermodynamic entropy}---Finally, we provide a geometrical interpretation of the thermodynamic model described above in terms of the random-chain polymer ensemble \cite{Gies06}; this approach was inspired by the worldline formalism \cite{Schubert01}. We first note that the functional integral of the gradient term $\int D\theta \exp\left[-\int (\nabla \theta)^2\right]$ has an entropic origin as it sums over configurations of phantom chain polymers in free space \footnote{The mapping between phantom or self-avoiding polymers and field theory is discussed in PG de Gennes, \emph{Scaling concepts in polymer physics}, Cornell university press, 1979. See also \cite{Gies06}.}. In the presence of boundaries [Eq.~(\ref{Eq. B.C.})], this sum is further weighted when a polymer intersects a boundary.
In fact, to compute $\Delta F_{A,B}(\lambda)$, one must sum over chain polymers that intersect both $A$ and $B$ regions, weighted by their size and intersections, the latter through $\lambda$ (see Fig. 1).
In this sense, the change in free energy is nothing but the change of the entropy ($k_B T=1$), that is $-\Delta F =\Delta{\cal S}_{\rm th}$ where ${\cal S}_{\rm th}$ denotes the purely thermodynamic entropy of a fluctuating polymer.
We thus arrive at the fundamental conclusion of this Letter: The QMI in $D$ dimensions is obtained from the \emph{thermodynamic entropy} in $D+1$ dimensions as
 \begin{equation}\label{Eq. Thermodynamic dictionary}
   {\cal I}^{(D)}=2\omega_c \int_{0}^{\infty} \!\!\!d\lambda \,\, \Delta {\cal S}^{(D+1)}_{\rm th}(\lambda),
 \end{equation}
where we dropped the dependence on $A,B$, but made the dependence on the dimension explicit.
We stress that Eq.~(\ref{Eq. Thermodynamic dictionary}) and the above geometrical picture is easily generalized to three or more regions. For example, the quantum tripartite information is due to polymer chains that intersect all three regions. In fact, the strong subadditivity property for quantum systems \cite{Lieb73} expressed in terms of mutual and tripartite information becomes rather trivial once formulated in the geometrical picture above; see the {Supplemental Material} for details. The thermodynamic analogy (sum over classical configurations) is crucial here.

\emph{Discussion and outlook}---We have studied the entanglement of
dispersive media, and shown that their mutual (and tripartite, etc.) information maps to the
thermodynamic entropy in a higher dimension. Both analytical and numerical
computations are illustrated based on an electrostatic analogy.
Extending the results of this paper to realistic models such as
electromagnetism is worthwhile. The response function can also be
generalized to a sum of Lorentzians, and thus a general $\epsilon(\omega)$,
at the expense of introducing several copies of the matter field for
each term in the sum. Including finite temperature should be of interest from both
fundamental and practical perspectives. Finally, lifting the assumption
of weak coupling as well as generalizing to non-linear and conformal
field theories are worthwhile but more challenging.
\\

{\bf Acknowledgements:} We thank Frank Wilczek for many stimulating discussions specially regarding the strong subadditivity property.
 We acknowledge useful discussions with R.L. Jaffe, M. Kardar, A. Gorshkov, M. Hertzberg, B. Swingle, E. Tonni, and J. Sonner. This work was supported by NSF PFC at JQI, NSF PIF, ARO, ARL, AFOSR, and AFOSR MURI.

\begin{widetext}
\newpage
\begin{center}
  {\Large {\bf Supplemental Material}}
\end{center}
\setcounter{equation}{0}
\renewcommand{\theequation}{S\arabic{equation}}
\setcounter{figure}{0}
\renewcommand{\thefigure}{S\arabic{figure}}

\maketitle

  \section{The covariance matrix $\D$}

    Computing the covariance matrix $\D$ requires two-point functions. The latter should be computed by inverting the operator (cf. Eq.~(3) of the main text)
    \begin{equation}
      \frac{\xi^2+\omega_0^2}{4\pi}\,  \delta(\bx-\bx')+\omega_p^2\xi^2 G_\xi(\bx,\bx')
    \end{equation}
    defined on the support of the media $V$. To invert this operator, we will treat the second term perturbatively justified in the weak-coupling limit where $\omega_{0,p} L \ll 1$ with $L$ the linear size of the media. Therefore, to the first order, we find
    \begin{align}\label{Eq. psi-psi correlation}
      \langle \psi_\xi (\bx)\psi_\xi (\bx')\rangle
      &\approx \frac{4\pi}{\xi^2+\omega_0^2}\,\delta(\bx-\bx')-\left(\frac{4\pi\omega_p \xi}{\xi^2+\omega_0^2}\right)^2G_\xi(\bx,\bx').
    \end{align}

    The two-point functions $\bQ$ and $\bP$ defined in the manuscript can be computed accordingly. The field correlation function reads
     \begin{equation}
      \bQ(\bx,\bx')= 2\int_0^\infty \frac{d\xi}{2\pi} \langle \psi_\xi (\bx)\psi_\xi (\bx')\rangle\approx\frac{2\pi}{\omega_0}\delta(\bx-\bx') - 16 \pi\omega_p^2\int_0^\infty {d\xi} \frac{\xi^2}{(\xi^2+\omega_0^2)^2} G_\xi(\bx,\bx').
      \end{equation}
    where the factor of 2 in the first equality comes from the restriction to $\xi \in [0,\infty)$.
    Similarly, the correlator of the conjugate momentum is given by
    \begin{equation}
      \bP(\bx,\bx')=-2\int_0^\infty \frac{d\xi}{2\pi} \xi^2\langle \psi_\xi (\bx)\psi_\xi (\bx')\rangle
      \approx \frac{\omega_0}{8\pi}\delta(\bx-\bx') +\frac{\omega_p^2}{\pi}\int_0^\infty d\xi \frac{\xi^4}{(\xi^2+\omega_0^2)^2} G_\xi(\bx,\bx'),
    \end{equation}
    with the minus sign in the first equality due to $\omega \to i \xi$. Superficially, the integral over the first term in Eq.~(\ref{Eq. psi-psi correlation}) is divergent; however, a proper regularization is to compute $\lim_{t\to 0^+}\langle p(t) p(0)\rangle$ ($p$ representing the conjugate momentum) which yields the first term in the last equality above \cite{SWeiss99}. The last term in this equality can be simplified by dropping $\omega_0$ in the denominator as $\omega_0 L \ll 1$. It is useful to note that
    \begin{equation}
      \int_{-\infty}^{\infty}\frac{d\omega}{2\pi} \,G_\omega^{(D)}(\bx)=G_0^{(D+1)}(\bx,0),
    \end{equation}
    where the superscript denotes the dimension, $(\bx, x_{D+1})$ represents coordinates in one higher dimension, and $G_0$ is the Green's function corresponding to the Laplacian. Defining $\delta \bQ$ and $\delta \bP$ as in the manuscript, the above equations yield $\delta \bP(\bx,\bx') \approx (4\pi \omega_p^2/\omega_0)G_0^{(D+1)}(\tilde\bx,\tilde\bx')$ with $\tilde \bx=(\bx,0)$, while $\delta \bQ$ can be neglected compared to $\delta \bP$. With $\D\approx (\delta\bQ+\delta\bP)/2$ (cf. the manuscript), we recover the expression for $\D(\bx,\bx')$ in Eq.~(5) of the manuscript.

  \section{The capacitance of a disc in $D+1=3$ dimensions}
  In this section, we consider a modified electrostatic problem in 3 dimensions. The `potential' $\theta(x,y,z)$ satisfies the Laplace equation, $\nabla^2 \theta=0$, except on the surface of a disc of radius $R$ sitting at the $z=0$ plane. The disk imposes the $\lambda$-dependent boundary condition (9) on $\theta$ (whose dependence on $\lambda$ is implicit). It is convenient to consider the oblate spheroidal coordinates \cite{SMorse53},
  \begin{align}
    &x= R \, \sqrt{(\xi^2+1)(1-\eta^2)} \cos \phi,\nonumber\\
    &y= R \, \sqrt{(\xi^2+1)(1-\eta^2)} \sin \phi,\\
    &z= R \, \xi \eta, \nonumber
  \end{align}
  where $0\le \xi <\infty$, $-1 \le \eta \le 1$, and $0\le\phi<2\pi$.
  With the reflection symmetry $z \to -z$, we separately consider even and odd solutions with respect to the $z$ coordinate. Odd solutions become trivial as $\theta(x,y, z=0)=0$ whereby the boundary condition simply requires the continuity of the first derivative. For even solutions, the boundary condition becomes $-2\partial_n \theta+{\lambda}^{-1}\,\theta(x,y,z=0)=0$ where $\partial_n$ is the normal derivative on the surface. In oblate spherical coordinates, the latter condition takes the form
  \begin{equation}
    \Big[-\partial_\xi \theta+\frac{R}{2\lambda} |\eta| \,\theta\,\Big]_{\xi=0}=0.
  \end{equation}
  The regular solutions (decaying at infinity as $\xi\to \infty$) of the Laplace equation are given by
  \begin{equation}
    {\mathcal Q}_n(i\xi) P_n^m(\eta) e^{i m \phi},
  \end{equation}
  where
  \(
  {\cal Q}_n(i x)=   Q_n(i x)-\frac{i \pi}{2} P_n(ix)
  \)
  with $P$ ($Q$) the Legendre polynomial of the first (second) kind.
  To find the capacitance, we consider a constant potential of magnitude 1 at $\xi \to \infty$, to which the object responds in a multi-pole series expansion as
  \begin{equation}\label{Eq: chi}
    \theta(\xi,\eta,\phi)= 1+ c_0 {\cal Q}_0(i\xi) +\sum_{{\rm Even\, }n\ge 2}c_n {\cal Q}_n(i\xi) P_n(\eta).
  \end{equation}
  We have exploited the symmetries of a disc to restrict the sum to even $n$, and choose $m=0$. The boundary condition then takes the form
  \begin{equation}\label{Eq: BC expansion}
    -c_0 {\cal Q}_0'-\sum_{{\rm Even\, }n\ge 2}c_n {\cal Q}_n' P_n(\eta)+\frac{R}{2\lambda} |\eta|\left(1+c_0 {\cal Q}_0+\sum_{{\rm Even\, }n\ge 2}c_n {\cal Q}_n P_n(\eta)\right)=0,
  \end{equation}
  where we have defined
  \(
    {\cal Q}_n(0)={\cal Q}_n
  \)
  and
  \(
  d{\cal Q}_n(i\xi)/d\xi\big|_{\xi=0}={\cal Q}_n'.
  \)
  Integrating over $\eta$, Eq.~(\ref{Eq: BC expansion}) yields
  \begin{equation}
    -c_0{\cal Q}_0'+\frac{R}{4\lambda}\left(1+c_0 {\cal Q}_0+\sum_{{\rm Even\, }n\ge 2} c_n {\cal Q}_n\int d\eta\, |\eta| P_n(\eta)\right)=0.
  \end{equation}
  Comparing with numerical evaluation, one can see that ignoring higher multipoles, i.e. dropping the sum over $n>2$, is an excellent approximation (and, at least, exact in the two limits where $R/\lambda \ll 1$ and $R/\lambda\gg 1$); we can then solve for the lowest multipole coefficient (using ${\cal Q}_0=-i\pi/2$ and ${\cal Q}_0'=i$),
  \begin{equation}
    c_0\approx\frac{-i}{4\lambda /R +\pi/2}.
  \end{equation}
  Inserting $c_0$ in Eq.~(\ref{Eq: chi}), we find that, at large radius $r=\sqrt{x^2+y^2+z^2}$, the potential behaves as
  \(
    \theta\sim 1-\frac{i R c_0}{r}
  \)
  (using $Q_0(i\xi) \sim -i/\xi$ and $\xi\sim r /R$ for large $\xi$).
  The capacitance is then given by
  \(
    C=i R c_0\approx \frac{R}{4\lambda /R +\pi/2}.
  \)
  This formula correctly reproduces the Dirichlet case for $\lambda=0$ \cite{SMorse53}.

 \section{Interpretation in terms of random chain polymers}
  In this section, we outline an alternative approach to computing the free energy of a thermally fluctuating field based on the worldline formalism \cite{SGies06}; this technique was first applied to computing the Casimir interaction energy between two objects. We shall consider the objects as infinitely thin surfaces represented by a $\delta$-function potential [$\tilde \bx=(\bx,x_{D+1})$ denotes the coordinates in $D+1$ dimensions]
  \begin{equation}
    V_A(\tilde \bx)=\lambda^{-1}\int_A d\Sigma  \, \delta (\tilde \bx-\tilde \bx_\Sigma),
  \end{equation}
  defined on a $D$ dimensional hyper-surface $A$ in $D+1$ dimensions. Note that the limit $\lambda \to 0$ corresponds to Dirichlet boundary conditions. Now consider a worldline $\tilde \bx(\tau)$ parameterized by the variable $\tau$ which is normalized as the Euclidean length of the worldline trajectory, and define
  \begin{equation}
    {\cal T}_{A}[\tilde \bx(\tau)]=\int d\tau \, V_A(\tilde \bx(\tau)).
  \end{equation}
  We are interested in finding the thermodynamic entropy in the presence of two hyper-surfaces $A$ and $B$. The change of the thermodynamic entropy $\Delta {\cal S}_{\rm th}$ (cf. the manuscript for the definition) can be expressed as a sum over worldline loops, or closed phantom polymers, as
  \begin{equation}\label{Eq. polymer two objects}
    \Delta {\cal S}_{\rm th}(A,B;\lambda)\propto\int \frac{dl}{l^{1+\frac{D+1}{2}}}\int d\tilde \bx_{\rm CM} \, \left\langle 1-e^{-{\cal T}_{A }}-e^{-{\cal T}_{ B}}+e^{-{\cal T}_{A \cup B}}   \right\rangle,
  \end{equation}
  where $l$ and $\tilde \bx_{\rm CM}$ are the length and the center of mass of a worldline trajectory, respectively, and the average $\langle \cdot \rangle$ is taken over all loops with a fixed $l$ and $\tilde \bx_{\rm CM}$; see Ref.~\cite{SGies06} for more details. We also made explicit the dependence of thermodynamic entropy on the hyper-surfaces $A$ and $B$. Finally, the proportionality coefficient is a $D$-dependent constant, which will not be important for our purposes. The above equation has a simple interpretation:
  Only worldlines that ``see'' both surfaces contribute to $\Delta {\cal S}_{\rm th}$. For the special case of Dirichlet boundary conditions, $\lambda=0$, the integrand is 1 when the worldline intersects both $A$ and $B$, and 0 otherwise, since
  \begin{equation}\label{Eq. lambda=0}
    e^{-{\cal T}_{A}}=
    \begin{cases}
      1, & \mbox{polymer does not intersect $A$} \\
      0, & \mbox{polymer intersects $A$}
    \end{cases}
  \end{equation}

 \section{Quantum tripartite information}
  In the main text, we derived an expression for the mutual information between two material bodies in terms of the thermodynamic free energy, and consequently the thermodynamic entropy, in one higher dimension. Here, we generalize this result to three bodies. Further generalization to more than three regions is straightforward.
  First we define the quantum tripartite information ${\cal I}(A,B,C)$ for three systems $A$, $B$ and $C$ as \cite{SCasini09}
  \begin{equation}\label{Eq: entanglement 3 objs}
    {\cal I}(A,B,C)={\cal S}(A)+{\cal S}(B)+{\cal S}(C)-{\cal S}(A\cup B)-{\cal S}(A\cup C)-{\cal S}(B\cup C)+{\cal S}(A\cup B\cup C),
  \end{equation}
  where $\cal S$ represents the von-Neumann entropy. Again this definition is free of UV divergences.
  Therefore, the tripartite quantum information maps to the thermodynamic entropy as (also making the dependence on the dimension $D$ explicit)
  \begin{equation}\label{Eq. Thermo analogy three objects}
    {\cal I}^{(D)}(A,B,C)=2\omega_c\int_{0}^{\infty} d\lambda \, \Delta_3{\cal S}^{(D+1)}_{\rm th}(A,B,C;\lambda).
  \end{equation}
  where we have defined (making the dependence on $\lambda$ implicit) $\Delta_3 {\cal S}_{\rm th}(A,B,C)\equiv {\cal S}_{\rm th}(A)+{\cal S}_{\rm th}(B)+{\cal S}_{\rm th}(C)-{\cal S}_{\rm th}(A\cup B)-{\cal S}_{\rm th}(A\cup C)-{\cal S}_{\rm th}(B\cup C)+{\cal S}_{\rm th}(A\cup B\cup C)$.
  A trace expression in terms of capacitance matrices similar to Eq.~(8) of the main text, but for three objects, can be found in Ref.~\cite{SEmig08}.

 \section{Strong subadditivity property}
 For any tripartite quantum system comprising disjoint subsystems $A, B,$ and $C$, the strong subadditivity theorem states that \cite{SNielsen10}
  \begin{equation}\label{Eq: strong subadditivity}
   {\cal S}(A\cup B\cup C)+{\cal S}(B) \le {\cal S}(A\cup B) + {\cal S}(B\cup C).
  \end{equation}
  This inequality can be recast in terms of the mutual and tripartite information (cf. the manuscript and the previous section for the definitions) as
  \begin{equation}
    {\cal I}(A,B,C)\le {\cal I}(A,C) .
  \end{equation}
  The proof of the above theorem is highly nontrivial. Nevertheless, using the thermodynamic analogy, Eq.~(\ref{Eq. Thermo analogy three objects}), the subadditivity property can be easily verified in our model. More precisely, we show that
  \begin{equation}\label{Eq. Strong subadditivty for S-th}
    \Delta_3{\cal S}_{\rm th}(A,B,C;\lambda)\le \Delta{\cal S}_{\rm th}(A,C;\lambda)
  \end{equation}
  for each value of $\lambda$. The quantum strong subadditivity property follows immediately.
  To this end, we cast the change of the thermodynamic entropy for three objects in terms of polymer chains:
  \begin{align}
    \Delta {\cal S}_{\rm th}(A,B,C;\lambda)\propto
    \int \frac{dl}{l^{1+\frac{D+1}{2}}}\int d\tilde \bx_{\rm CM} \, \left\langle 1- e^{-{\cal T}_{A }}-e^{-{\cal T}_{B}}-e^{-{\cal T}_{ C}}
    +e^{-{\cal T}_{A \cup B}}+ e^{-{\cal T}_{A \cup C}} + e^{-{\cal T}_{B\cup C}}
    -e^{-{\cal T}_{A \cup B\cup C}}\right\rangle.
  \end{align}
  Similar to the mutual information, only worldlines that see all three surfaces contribute to $\Delta_3{\cal S}_{\rm th}$. Equation (\ref{Eq. Strong subadditivty for S-th}) then finds a pictorial meaning: The contribution of polymer configurations that intersect all three regions $A$, $B$ and $C$ is less than or equal to those that intersect $A$ and $C$, but not necessarily $B$. In fact, for the special case of $\lambda=0$, the inequality in Eq.~(\ref{Eq. Strong subadditivty for S-th}) becomes rather trivial as it simply means [cf. Eq.~(\ref{Eq. lambda=0})] that the number of polymers that intersect all three regions is less than or equal to the number of those which intersect any two regions, but not necessarily the third one.
  To prove the inequality in general, we define
  \begin{align}
    e^{-{\cal T}_{A}}=x,\quad  e^{-{\cal T}_{B}}=y,\quad  e^{-{\cal T}_{C}}=z,
  \end{align}
  hence,
  \(
  e^{-{\cal T}_{A\cup B}}=x y
  \)
  and
  \(
  e^{-{\cal T}_{A\cup B\cup C}}=x y z,
  \)
  etc.
  To derive Eq.~(\ref{Eq. Strong subadditivty for S-th}), it is sufficient to show that
  \begin{equation}
     1-x-y-z+xy+yz+xz -xyz \le 1-x-z+xz
  \end{equation}
  but this is simply equivalent to $(-1+x)y(-1+z) \ge 0$ which is trivially satisfied (note that $0\le x,y,z \le 1$).

\section{Numerical computations}

As noted in the main text, an advantage of our method is that it
expresses quantum mutual information in terms of classical scattering
data, thus opening the door to myriad existing methods for practical
computations. Indeed, in principle any method of computational
electrostatics could be used to calculate the quantities
$\vb C(\lambda)$ and $\vb G_0$ that enter Eq.~(8) of the main text;
this includes both analytical techniques such as multipole
expansions and numerical methods such as finite-difference
or boundary-element approaches.

The choice of a computational method may be viewed as
the choice of a discrete \textit{basis} in which to
evaluate matrix elements of the $\vb C(\lambda)$ and
$\vb G_0$ operators in Eq.~(8) of the main text. Many different choices of
basis are possible; we here consider two in particular:
\begin{enumerate}
 \item For highly symmetric shapes at medium or long distances,
       it is convenient to introduce a multipole basis
       that takes advantage of symmetries to encode scattering
       data in compact form. In some cases the representation is
       so concise that Eq.~(8) of the main text may be evaluated analytically,
       as in the case of the discs at large separation discussed in the main text.
 \item For more general shapes, or in the short-distance limit,
       it is convenient to introduce a basis of localized functions
       supported only on the material bodies.
       This allows treatment of essentially arbitrary shapes---such
       as the squares of Figure 2, for which a multipole expansion
       would be cumbersome---and calculations at essentially
       arbitrarily short distances.
\end{enumerate}

Before describing each of these possibilities in turn, we pause
to note that this dichotomy mirrors the recent evolution of theoretical
approaches to electromagnetic fluctuation phenomena (Casimir forces and
near-field radiative heat transfer), in which geometry-specific
multipole bases have been employed to yield rapidly convergent
or even analytically tractable series expansions of Casimir
quantities and heat-transfer rates in high-symmetry
geometries~\cite{SRahi09, SKrueger2012}, while localized basis
functions have been used for numerical Casimir or heat-transfer
calculations in arbitrary
geometries~\cite{SReid2013B, SRodriguez2013B}.

\subsection*{Multipole basis}

For objects of high symmetry---such as the disc-shaped
material bodies considered in the main text---it is convenient to
work in a multipole basis. For example, in a basis of
spherical multipoles one expands scalar fields in solutions
of the Laplace equation indexed by the usual spherical indices
$\ell,m$:
\begin{equation}
\phi\sups{int}_{\ell m}(\vb r)
   = r^\ell Y_{\ell m}(\theta,\phi),
   \qquad
   \phi\sups{ext}_{\ell m}(\vb r)
   = \frac{1}{r^{\ell + 1}} Y_{\ell m}(\theta,\phi)
\label{PhiIntExt}
\end{equation}
where $Y_{\ell m}$ are the usual spherical harmonics. In this case,
\begin{itemize}
\item
the $\vb G_0$ matrices in Eq.~(8) of the main text become translation matrices,
which relate spherical multipoles centered at the origin of
one body to multipoles centered at the origin of another body,
while
\item
the elements of the $\vb C(\lambda)$ matrices in Eq.~(8) of the main text describe the
multipole sources induced on
(or, equivalently, the outgoing multipole fields emitted by)
material bodies by an incident multipole field.
\end{itemize}
More specifically, the definition of the spherical-wave $\vb C$ matrix
elements for a body is as follows: If the body is exposed to an
``incident'' field consisting of a single incoming multipole field,
i.e.  $\phi\sups{inc}=\phi\sups{int}_{\ell, m}$, then the
``scattered'' field produced by the body is
$$ \phi\sups{scat}
   =\sum_{\ell^\prime, m^\prime}
    C_{\ell^\prime, m^\prime; \ell, m} \phi_{\ell^\prime, m^\prime}\sups{ext}(\vb r).
$$
Thus, to compute one full column of the $\vb C$-matrix we solve
a modified electrostatics problem in which the body [on whose
surface we have the $\lambda$-dependent boundary condition (7)]
is exposed to an external $(\ell,m)$ field; after solving the
scattering problem we extract
$C_{\ell^\prime, m^\prime; \ell, m}$
as the $(\ell^\prime,m^\prime)$ source
multipole induced by this field on the body.

\subsection*{Localized basis}

As noted in the main text, the multipole basis
becomes unwieldy in the short-distance limit due to the
need to retain large numbers of multipoles to obtain
sufficient accuracy. Moreover, even at long distances the
\textit{analytical} usefulness of the multipole basis
is restricted to highly symmetric objects
for which the $\vb C$-matrix elements may be computed
in closed form.
For shorter distances or less symmetric
objects (such as the squares of Figure 2), it is convenient
to work with position-space representations of the
$\vb C$ and $\vb G_0$ operators in Eq.~(8) of the main text.

A direct position-space interpretation of the operator
trace in Eq.~(8) of the main text is ill-defined for at least two reasons.
One difficulty is that the operator $\vb G_0$, whose
position-space representation is
\begin{equation}
 G_0(\vb x, \vb x^\prime) = \frac{1}{4\pi|\vb x-\vb x^\prime|}
 \label{G0PositionSpace}
\end{equation}
is singular ``on the diagonal,'' i.e. for $\vb x=\vb x^\prime.$
This singularity needs somehow to be regularized before
the operator products and trace in Eq.~(8) of the main text may be performed.
A second difficulty is that the $\vb C(\lambda)$ operators are unwieldy
in position space. Indeed, for e.g. body A the position-space
matrix elements $C_A(\lambda; \vb x, \vb x^\prime)$ satisfy the
following mixed definition:
\begin{enumerate}
 \item If either of the points $\vb x, \vb x^\prime$ lies
       outside of object $A$, then we have an explicit representation
       for $C_A(\lambda; \vb x, \vb x^\prime)$, namely
       $$C_A(\lambda; \vb x, \vb x^\prime) = 0
         \quad \text{ if } \quad
         \vb x \notin A
         \quad \text{ or } \quad
         \vb x^\prime \notin A.
       $$
 \item On the other hand, if both points $\vb x, \vb x^\prime$ lie
       inside body $A$ then we have only an explicit representation
       for the \textit{inverse} of the $\vb C$ operator,
       \begin{equation}
           C^{-1}_A(\lambda; \vb x, \vb x^\prime)
          = \lambda \delta(\vb x-\vb x^\prime)
            + \vb G_0(\vb x, \vb x^\prime).
         \label{CInvPositionSpace}
       \end{equation}
\end{enumerate}

Both of these difficulties are alleviated by introducing a finite
basis of $N_A$ localized functions
$\{b_{An}(\vb x)\}, n=1,\cdots,N_A,$
whose supports are restricted to a finite subregion of body $A.$
For example, we might discretize the body into $N_A$
small triangles (as depicted in the inset of Figure 2);
to the $n$th triangle $M_{An}$ we associate a basis function
$b_{An}(\vb x)$ defined to be unity for $\vb x\in M_{An}$
and vanishing elsewhere. [Similarly, for body $B$
we introduce $N_B$ basis functions $\{b_{Bn}(\vb x)\}$ supported
on subregions of body $B$.]

With this choice of basis functions, the elements of the
$\vb G$ operator take the form of four-dimensional
integrals over triangle-product domains:
\begin{align}
   G_{\alpha\beta}
    =\Big\langle b_{\alpha} \Big| \vb G \Big| b_{\beta} \Big\rangle
   &=\int \int \frac{b_{\alpha}(\vb x) b_{\beta}(\vb x^\prime)}
                    {4\pi|\vb x-\vb x^\prime|}\,d\vb x\,d\vb x^\prime
\nonumber\\
   &=\int_{M_{\alpha}}\int_{M_{\beta}}
     \frac{1}{4\pi|\vb x-\vb x^\prime|}\,d\vb x\,d\vb x^\prime.
\label{G0mn}
\end{align}
(Note that the $\vb G$ matrix in the localized basis
is nonsingular even on the diagonal,
i.e. $G_{\alpha\alpha}<\infty;$
the finite extent of the basis functions regularizes the
position-space divergence of equation (\ref{G0PositionSpace}).)
The $\vb G$ operators in Eq.~(8) of the main text become $N_A \times N_B$
or $N_B\times N_A$ matrices describing the interactions
of triangles on body $A$ with triangles on body $B$.

The $\vb C_{A}$ and $\vb C_{B}$ operators in Eq.~(8) of the main text become
$N_A\times N_A$ and $N_B\times N_B$ matrices. To compute
these it is easiest to use equation (\ref{CInvPositionSpace})
to compute first the inverse matrices $\vb C_{A,B}^{-1}$.
The elements of e.g. $\vb C_A^{-1}$ read
\begin{align}
 C^{-1}_{A;mn}(\lambda)
   &=\Big\langle b_{Am} \Big| \vb C^{-1}_A(\lambda) \Big| b_{An} \Big\rangle
\\
   &=\lambda \mathcal{N}_m \delta_{mn}
     + \Big\langle b_{Am} \Big| \vb G_0 \Big| b_{An} \Big\rangle
\end{align}
where $\mathcal{N}_m=\int b^2_m(\vb x)\,d\vb x$
is the norm of basis function $b_m$
and the $\vb G_0$ matrix element is computed as in (\ref{G0mn}).
Having computed the full numerical matrix $\vb C^{-1}(\lambda)$,
we then simply invert this matrix numerically to obtain
the $\vb C$ matrices that enter Eq.~(8) of the main text. Matrix products
are computed using standard numerical linear algebra software,
and the logarithm is computed using the identity
$ \text{Tr}\log \vb M = \sum \log \lambda_n$
where $\{\lambda_n\}$ are the eigenvalues of $\vb M$.
Finally, the $\lambda$ integral in Eq.~(8) of the main text is evaluated using
an adaptive numerical quadrature method.

\end{widetext}

\end{document}